# Contactless measurements of the elastic modulus of living cells using thermal fluctuations of atomic force microscope cantilever

Hao Zhang[1,2], Zaicheng Zhang[3], Etienne Harte[1], Francoise Argoul[1], Abdelhamid Maali1[1,*]

[1] *Université de Bordeaux & CNRS, LOMA, UMR 5798, F-33400 Talence, France*
[2]*Xiamen Institute of Rare Earth Materials, Haixi Institute, Chinese Academy of Sciences, 361021 Xiamen, China*
[3] *School of Physics, Beihang University 100191 Beijing, China*

**Abstract:** We present a contactless method for measuring the elastic modulus of living cells (human triple-negative breast cancer, MDA-MB-231) from the thermal fluctuations of an atomic force microscope (AFM) cantilever. By analyzing the power spectral density (PSD) of the cantilever's thermal fluctuations, we obtain the resonance frequencies of its first three modes at various cell to cantilever separation distances. By comparing measurements on living cells with those on a rigid borosilicate sphere of the same size, we extract the frequency shift caused by the elasto-hydrodynamic coupling between the cantilever fluctuations and the deformations of the cells. We then fit this frequency shift using an elasto-hydrodynamic model that integrates hydrodynamic forces and cell deformation. This approach allows us to determine the elastic modulus values of the living cells for the first three resonant frequencies of the cantilever.

\* Corresponding author:
Abdelhamid Maali: abdelhamid.maali@u-bordeaux.fr

**Introduction:**

The mechanical properties of living cells play a crucial role not only in fundamental cellular processes such as migration, adhesion, and growth but also in various pathological conditions such as cancers, myopathies, cardiovascular, and neuronal disorders. A wide range of methods, including magnetic twisting cytometry (MTC), particle tracking microrheology, optical stretching, parallel plate rheology, and indentation techniques such as atomic force microscopy (AFM) or cellular force microscopy (CFM), have been used to characterize the mechanical properties of single cells or cell layers [1-15]. However, many of these methods involve physical and intrusive contact with the cell, which can induce stretching, compression, or even disruption of the cell structure, often resulting in local or global fractures [5, 13, 16], the forces exerted during indentation techniques being large, typically ranging from nano-Newtons to hundreds of micro-Newtons.

Non-contact measurement techniques offer an alternative approach to overcome the limitations of direct-contact methods. In particular, the elastohydrodynamic (EHD) interaction between a living cell and an AFM probe, separated by a thin liquid film, has emerged as a promising method [17-23]. As a result, the liquid film acts as a coupling hydrodynamic medium, preventing direct contact with the cell. When the gap between the sphere and the cell surface is very small, the vibration of the sphere generates hydrodynamic pressure that can deform the cell, resulting in EHD coupling. Based on this coupling, several tools have been developed to probe the mechanical properties of samples of different stiffness without physical contact [17-23].

Our recent work has demonstrated the utility of AFM as a powerful tool for investigating EHD coupling and non-contact measurement of nanoscale rheological properties of soft samples [17, 22-23].

In this work, we present a contactless method for measuring the elastic modulus of living cells (human triple-negative breast cancer, MDA-MB-231) using the thermal fluctuations of an AFM cantilever. The cantilever fluctuations generate a hydrodynamic viscous pressure that deforms the cell, which in turn modifies the flow and thus the thermal fluctuations of the cantilever. This elasto-hydrodynamic coupling between the cell deformations and the cantilever fluctuations is estimated from the power spectral density (PSD) of the cantilever thermal fluctuations at various separation distances. The frequency shift due to the cell deformation extracted from the PSD is then fitted with a simplified elasto-hydrodynamic model. The elastic modulus values of the living cell (MDA-MB-231) are determined for the three first resonance frequencies of the cantilever.

**Materials and methods:**

Experiments were performed using a JPK BioAFM system (Bruker) equipped with a temperature control module. A tipless cantilever (HQ: NSC36, NanoAndMore) with a static stiffness of $k_c = 0.03 \pm 0.002\ N/m$ was held on a cantilever holder specifically designed for operation in a liquid medium. To control the distance between the cantilever and the sample, we used an integrated step motor for coarse adjustment and a piezo for fine adjustment. The null distance was defined as the position where the cantilever deflection changes, indicating contact. Thermal fluctuations of the cantilever were recorded at a sampling rate of 500 kHz using an analog-to-digital (A/D) acquisition card (PXIe-4481, National Instruments). The PSD of the cantilever thermal fluctuations was calculated from 500,000 points, corresponding to an acquisition time of $1\ s$.

The MDA-MB-231 cells used in this experiment are human triple-negative breast cancer (TNBC) cells. They were cultured in Dulbecco's Modified Eagle Medium high glucose (Gibco, 31966-021) supplemented with 10% Fetal Bovine Serum (FBS) and passaged twice a week. Petri dishes (FD35-100 fluorodish) were cleaned and rendered hydrophilic by15-minute of plasma cleaning. These dishes were then incubated for 4 hours at 37°C with 3 mL of a 0.5 g/L aqueous solution of polyethylene glycol (PEG, Mw=3350 g/mol, Sigma Aldrich). Polyethylene glycol coating was designed to prevent cell adhesion to the surface, and maintain cell sphericity during the 2-hour experiment. This effect was confirmed using lens-less phase microscopy (Iprasense Cytonote). After PEG deposition, the petri dish was rinsed twice with phosphate buffer saline (PBS, Sigma Aldrich). The cells were then passaged to achieve a concentration of 100,000 cells per milliliter. Next, 1 mL of cells (i.e., 100,000 cells) were seeded into the petri dish and placed in an incubator for 40 minutes. During this time, the cells were observed under a bench phase microscope to ensure they maintained a spherical shape and remained attached to the substrate.

Borosilicate glass spheres (Duke Sci. Corp., Palo Alto, CA) of the same size as the cells (diameter of $16 \pm 1\ \mu m$) were chosen as references in our measurements. The spheres were rinsed successively in pure water and ethanol and then fixed on glass cover slide using epoxy glue.

## *Quality Factor and Resonance Frequency of the Cantilever*.

The cantilever beam exhibits different eigenmodes of vibration and the equation of motion at each mode can be modeled as a simple harmonic oscillator:

$$m_n^* \ddot{z} + \gamma_n^\infty \dot{z} + k_n z = F_n^{noise} + F_{int} \quad (1)$$

where $m_n^*$, $k_n$, and $\gamma_n^\infty$ represent the effective mass, the stiffness, and the bulk damping coefficient of the cantilever oscillation at the mode $n$, respectively. $F_n^{noise}$ is the thermal noise force and $F_{int}$ represents the interaction force with the sample. $z$ is defined as the vertical position of the AFM cantilever. The stiffness of each mode $k_n$ is related to the static stiffness $k_c$ by $k_n = \frac{\alpha_n^4}{12} k_c$, where the dimensionless parameter $\alpha_n$ obeys the relation: $\cos\alpha_n \cosh\alpha_n + 1 = 0$ [24-25]. The first three values of $\alpha_n$ and $k_n$ for a rectangular cantilever beam are given in **Table 1**.

*Table 1: The first three values of $\alpha_n$ and $k_n$, where $k_c$ is the static stiffness of the cantilever.*

| $n$ | 1 | 2 | 3 |
|---|---|---|---|
| $\alpha_n$ | 1.875 | 4.694 | 7.855 |
| $k_n = \alpha_n^4 k_c/12$ | $1.03\ k_c$ | $40.46\ k_c$ | $317.25\ k_c$ |

Taking the Fourier transform of **Eq. (1)** we obtain:

$$(-m_n^* \omega^2 + j\omega \gamma_n^\infty + k_n)\tilde{z} = \tilde{F}_n^{noise} + \tilde{F}_{int} \quad (2)$$

where $\tilde{z}$, $\tilde{F}_n^{noise}$, and $\tilde{F}_{int}$ denote the Fourier form of the variables, and $j$ is the imaginary unit. The interaction force $\tilde{F}_{int}$ can be written as the sum of an elastic and dissipative components:

$$\tilde{F}_{int} = -\bigl(G'(\omega) + jG''(\omega)\bigr)\tilde{z} \quad (3)$$

where $G'$ and $G''$ represent the real and imaginary components of the complex mechanical impedance, respectively.

Inserting **Eq. (3)** in **Eq. (2)** we get:

$$\left(-m_n^*\omega^2 + j\omega\left(\gamma_n^\infty + \frac{G''}{\omega}\right) + k_n + G'\right)\tilde{z} = \tilde{F}_n^{noise} \quad (4)$$

Then, we have:

$$|\tilde{z}(\omega)|^2 = \frac{\left(|\tilde{F}_n^{noise}|/m_n^*\right)^2}{(\omega^2 - \omega_n^2)^2 + \left(\frac{\omega \omega_n}{Q_n}\right)^2} \tag{5}$$

where $\omega_n = \sqrt{(k_n + G')/m_n^*}$ is the resonance frequency and $Q_n = \frac{m_n^* \omega_n}{\gamma_n}$ is the quality factor for mode $n$, with $\gamma_n = \gamma_n^\infty + \frac{G''}{\omega_n}$.

Using the fluctuation dissipation theorem $|\tilde{F}_n^{noise}| = 2\gamma_n k_B T$, the one-sided power spectral density $S(\omega) \equiv 2\langle|\tilde{z}(\omega)|^2\rangle$ can be derived as [26]:

$$S(\omega) = \frac{4k_B T}{m_n^*} \frac{\frac{\omega_n}{Q_n}}{(\omega^2 - \omega_n^2)^2 + \left(\frac{\omega \omega_n}{Q_n}\right)^2} \tag{6}$$

The resonance frequency $\omega_n$ and the quality factor $Q_n$ depends on the interaction between the cantilever and the sample. The expressions of $\omega_n$ and $Q_n$ are given by:

$$\omega_n = \sqrt{\frac{k_n + G'}{m_n^*}} \cong \omega_n^\infty \left(1 + \frac{G'}{2k_n}\right) \tag{7a}$$

$$Q_n = \frac{m_n^* \omega_n}{\gamma_n^\infty + \frac{G''}{\omega_n}} \cong \frac{Q_\infty}{1 + \frac{G''}{\gamma_n^\infty \omega_n^\infty}} = \frac{Q_n^\infty}{1 + \frac{Q_n^\infty G''}{k_n}} \tag{7b}$$

$\omega_n^\infty$, $Q_n^\infty$ are the values of the resonances frequencies and quality factors of the mode $n$ far from the surface (without interaction with the sample), and they are given as: $\omega_n^\infty = \sqrt{k_n/m^*}$ and $Q_n^\infty = m_n^* \omega_n^\infty / \gamma_n^\infty = k_n/(\gamma_n^\infty \omega_n^\infty)$.

Without loss of generality, to extract the resonance frequency and quality factor from the measurements, the experimental power spectral densities can be fitted in the form of :

$$S(\omega) = \frac{G_n}{(\omega^2 - \omega_n^2)^2 + \frac{\omega^2 \omega_n^2}{Q_n^2}} + C \tag{8}$$

where $G_n$ is a fitting parameter for each oscillation mode $n$, containing the conversion factor in the experiment and $C$ is a constant allowing to take into account a potential white noise background.

The PSD of the cantilever's fluctuation measured far from the sample is shown in **FIG. 1**. It shows the spectrum of the first three thermal vibration mode of the cantilever measured in the working liquid. The resonance frequency and quality factor for each mode are determined by fitting the resonance peaks individually using **Eq. (8)**. The red continuous line presents the fitting curve of the second mode of the cantilever vibration.

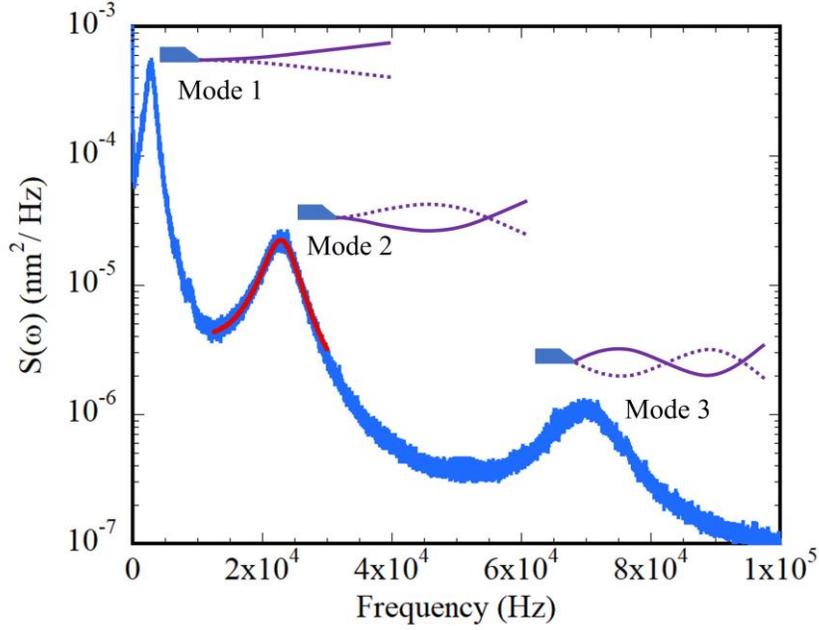

*FIG. 1: The thermal spectrum of a cantilever immersed in the working liquid far from the surface. Three distinct resonance peaks corresponding to the first three modes of the cantilever vibrations n=1, 2, 3 are clearly observed. The red continuous line represents an example of the fitting curve of the second resonance peak using **Eq. (8)**.*

From the measured PSD far from the sample, the fitted values of the bulk quality factors $Q_n^\infty$ and the bulk resonance frequencies $\omega_n^\infty$ for the first three modes of the cantilever are shown in **Table 2**.

Table 2: measured bulk quality factors and resonance frequencies for the first three modes of the cantilever.

| $n$ | 1 | 2 | 3 |
|---|---|---|---|
| $\omega_n^\infty/2\pi$ (Hz) | 3025 | 22938 | 69379 |
| $Q_n^\infty$ | 1.98 | 3.67 | 5.48 |

## Hydrodynamic interaction with a rigid sphere:

A cantilever oscillating in liquid near a rigid spherical object experiences a visco-elastic hydrodynamic force (see **FIG. 2**). The viscous component of the force is induced by the drainage of the liquid confined between the cantilever beam and the sphere and it is proportional to the instantaneous relative velocity between these two confining surfaces. The expression of this viscous component in the framework of the lubrication approximation is given by the Reynolds formula: $-\frac{6\pi\eta R^2}{d}\dot{z}$, where $\eta$ is the viscosity of the fluid, $R$ is the radius of the rigid sphere, $d$ is the separation distance between the cantilever and the sphere, and $\dot{z}$ represents the velocity of the oscillation.

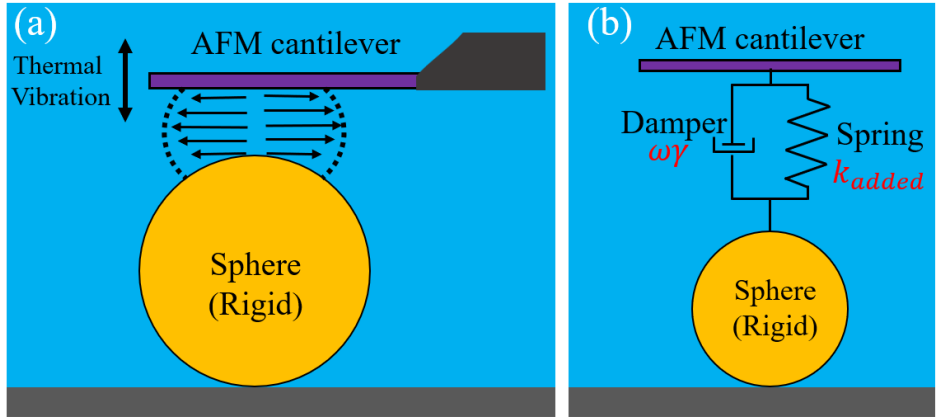

*FIG. 2 (a) Schematic diagram of the hydrodynamic interaction measurement between a tipless rectangular AFM cantilever and a rigid sphere. The distance between the cantilever and the sample is controlled by the AFM motor. (b) shows the equivalent model of the viscoelastic response of the rigid sphere, with a spring and dashpot in parallel.*

The elastic component is due to the surrounding liquid (added mass) oscillating with the cantilever near the sphere. Assuming that the hydrodynamic force acting on oscillating cantilever near a sphere is the same as the force acting on a cantilever at rest near an oscillating sphere, we can use the expression for the elastic force derived by *Fouxon* et al [27] for a sphere oscillating close to the wall given as: $-\frac{A\omega}{4(d+R)^3}z$. The parameter $A$ depends on the values of the frequency, the radius of the sphere, the viscosity and the density of fluid. Note that even though the above expression is derived assuming a large separation between the sphere and the surface, it is still valid at small separations [28-29]. In our case for each mode of the cantilever oscillation, we treat the value of $A$ as a fitting parameter.

The total interaction hydrodynamic force acting on the cantilever can be written as:

$$F_{int} = -k_{added}z - \gamma\dot{z} \tag{9a}$$

$$\gamma = \frac{6\pi\eta R^2}{d} \tag{9b}$$

$$k_{added} = \frac{A\omega}{4(d+R)^3} \tag{9c}$$

where $k_{added}$ is the effective elastic stiffness due to the added mass. These two contributions can be modeled as a damper and a spring in parallel as shown in **FIG. 2b**. The mechanical impedance of the interaction between the cantilever and the rigid sphere is then characterized by:

$$G_{rs} = -\frac{\tilde{F}_{int}}{\tilde{z}} = k_{added} + j\omega\gamma = \frac{A\omega}{4(d+R)^3} + j\frac{6\pi\eta R^2\omega}{d} \tag{10}$$

By inserting **Eq. (10)** into **Eqs. (7a)** and **(7b)**, we derive for each mode $n$ the expressions for the quality factor $Q_n^{rs}$ and the resonance frequency $\omega_n^{rs}$ as a function of the distance $d$:

$$Q_n^{rs}(d) \cong \frac{Q_n^{\infty}}{1 + \frac{6\pi\eta R^2 Q_n^{\infty}\omega_n^{\infty}}{k_n d}} \tag{11a}$$

$$\omega_n^{rs}(d) \cong \omega_n^{\infty}\left(1 + \frac{k_{added}}{2k_n}\right) \cong \omega_n^{\infty}\left(1 + \frac{A\omega_n^{\infty}}{8k_n(d+R)^3}\right) \tag{11b}$$

Given that in **Eq. (11a)**, all the values of the parameters are already determined, we can calculate $Q_n^{rs}(d)$ directly.

**FIG. 3a** shows the quality factor $Q_n^{rs}$ plotted as a function of distance d for the first three modes. The plain lines represent theoretical calculations based on **Eq. (11a)**, without the need for fitting parameters, since all physical parameter values in **Eq. (11a)** are known. Thus, the bulk quality factors $Q_n^{\infty}$ and the bulk resonance frequencies $\omega_n^{\infty}$ are shown in **Table 1**. The results show a good agreement between experimental measurements and theoretical calculations. In **FIG. 3b**, the frequencies are plotted against the distance for the first three resonances. The plain lines represent the fit curve using **Eq. (11b)**, with a single fitting parameter $A$ for each curve. The fitted values of A are -0.0016, -0.0051, and -0.0081 for mode 1, mode 2 and mode 3, respectively.

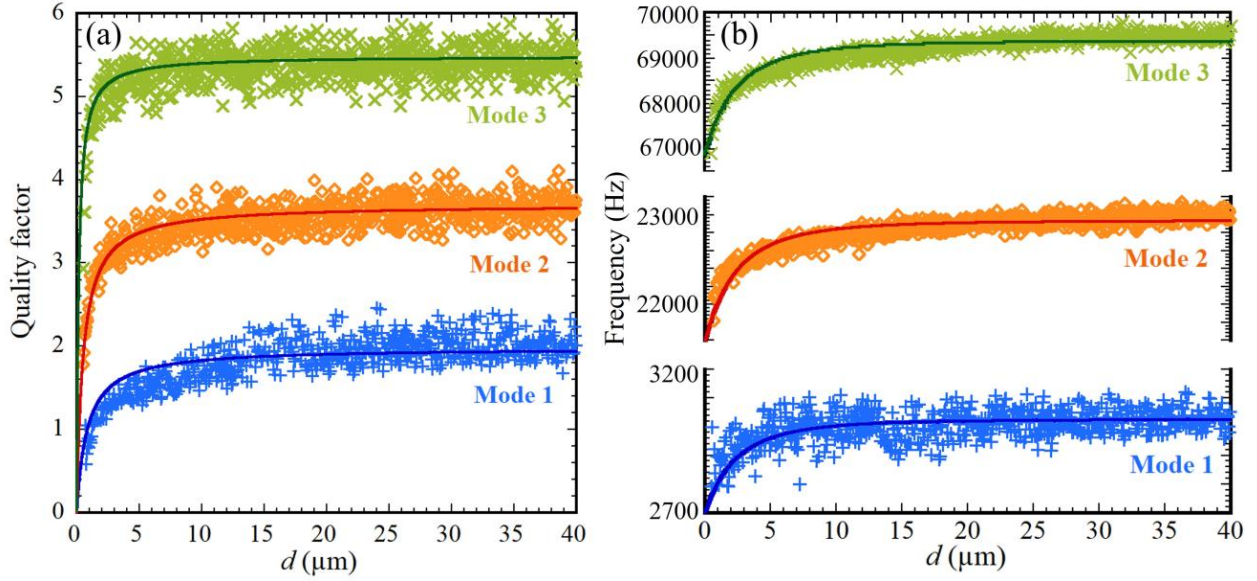

*FIG. 3 The quality factors (a) and resonance frequencies (b) versus the distance for the first three oscillation modes of the cantilever measured on the rigid sphere. The experimental data points are represented by dots. In (a), the plain lines represent calculations using **Eq. (11a)**, while in (b), the plain lines represent fitting results obtained using **Eq. (11b)**. The fitted values of A are -0.0016, -0.0051, and -0.0081 for mode 1, mode 2 and mode 3, respectively.*

## *Interaction with the cell:*

When the cantilever oscillates near a soft object such as cells, additional effects must be considered. The vertical motion of the cantilever generates a viscous hydrodynamic pressure that induces deformation of the living cells (see **FIG. 4a**), which in turn modifies the viscous flow and thus the motion of the cantilever near the cell. The elasto-hydrodynamic effect generated by the interplay between the hydrodynamic pressure and the deformation of the cell can be modeled using the Maxwell model (see **FIG. 4b**).

In our experiment, the nanoscale cantilever oscillation induces a very small deformation of the cell, allowing it to primarily probe the surface properties of the cell. We model the mechanical response of the cell by a spring of constant stiffness given as [19]:

$$K = \frac{4}{3}\pi E^{cell}\sqrt{2Rd} \tag{12}$$

where $R$ is the cell radius, $d$ the distance between the cantilever and the cell. Note here that we have considered a non-compressible cell, i.e. with a Poisson ratio of 1/2. $E^{cell}$ is the cell Young's modulus.

To obtain the total mechanical impedance of the system $(G_{tot}' + jG_{tot}'')$, we add the mechanical impedance of the cell in series with the impedance composed by the damper and the spring due to the added mass as previously calculated (interaction with rigid sphere).

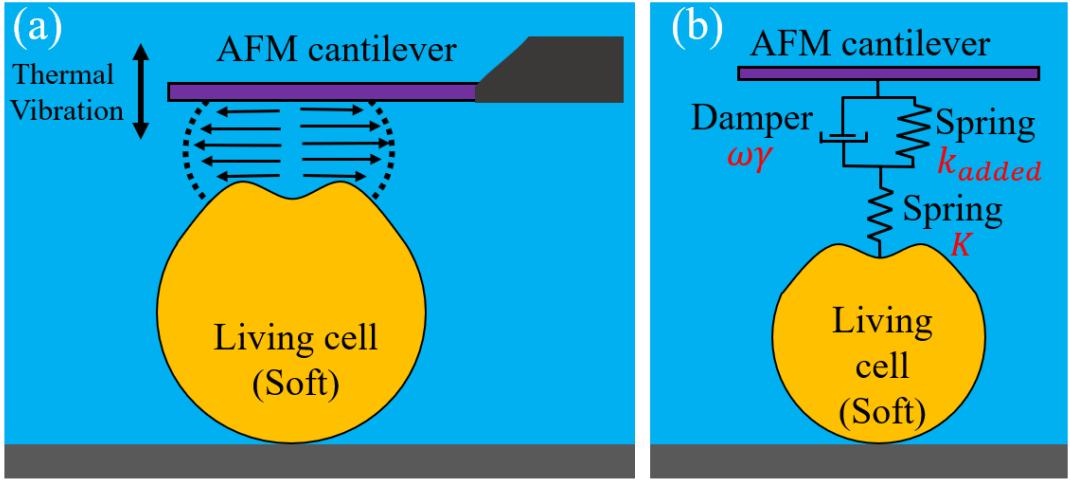

*FIG. 4 Schematic diagram of the hydrodynamic interaction measurement between a tipless rectangle AFM cantilever and a soft living cell. The hydrodynamic pressure generated by the vertical motion of the cantilever induces deformation of the living cells. b) show the equivalent model of the viscoelastic response of the living cell.*

The total mechanical impedance reads:

$$\frac{1}{G_{tot}' + jG_{tot}''} = \frac{1}{K} + \frac{1}{k_{added} + j\omega\gamma} \quad (13)$$

Under the condition of $k_{added} \ll \omega\gamma \ll K$, the leading order the elastic and inelastic part of the total mechanical impedance can be obtained as follows:

$$G_{tot}' \approx \frac{(\omega\gamma)^2}{K} + k_{added} \quad (14a)$$

$$G_{tot}'' \approx \omega\gamma \quad (14b)$$

Substituting **Eq. (14a)** into **Eq. (7a),** gives the resonant frequency for each oscillation mode $n$ as a function of the distance between the cantilever and the cell:

$$\omega_n^{cell} = \omega_n^\infty \left(1 + \frac{k_{added}}{2k_n} + \frac{(\omega\gamma)^2}{2k_n K}\right) \cong \omega_n^{rs} + \omega_n^\infty \frac{(\omega\gamma)^2}{2k_n K} \tag{15}$$

By subtracting **Eq. (11b)** from **Eq. (15)**, we obtain the expression for the frequency shift ($\delta\omega_n^{cell}$) induced by the EHD interaction between the cantilever fluctuations and the cell deformations for each oscillation mode $n$:

$$\delta\omega_n^{cell} = \omega_n^{cell} - \omega_n^{rs} = \frac{27\sqrt{2}\pi}{4} \frac{\eta^2 R^{\frac{7}{2}} \omega_n^{\infty 3}}{k_n E^{cell} d^{\frac{5}{2}}} \tag{16}$$

We can also calculate the frequency shift using the expressions derived by *S. Leroy* and *E. Charlaix* [19]. From the reported expression of the mechanical response at large distances, we obtain an expression very close to **Eq. (16)**.

In our experiments, the measurements on the rigid sphere serve as a reference. From the measured PSD, we extract the resonance frequency and the quality factor of the cantilever fluctuations as a function of the distance between the cantilever and the cell. **FIG. 5** shows the difference in the quality factor of the cantilever oscillating near the cell and near the rigid sphere. It is worth noting that the difference is approximately zero, and the sensitivity of our measurements is not sufficient to discriminate between the measured quality factor in two cases.

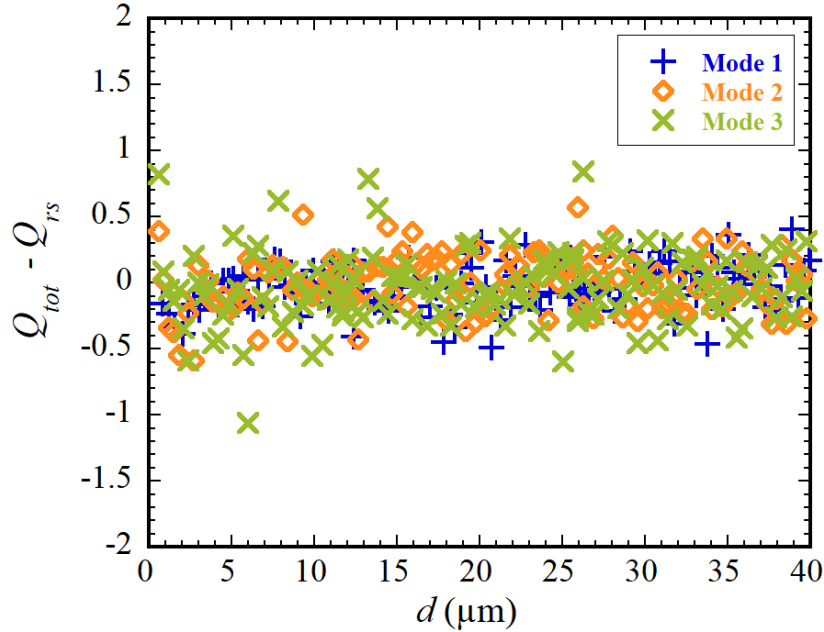

*FIG. 5 The difference in the value of the quality factor measured on the cell and rigid sphere as a function of distance for the first three modes.*

However, when the measured frequency obtained on rigid sphere is subtracted, the estimated frequency on the cell is not equal to zero as shown in **FIG. 6**. The frequency difference between the two cases is due to the elasto-hydrodynamic interaction. **FIG. 6** shows the frequency shift versus the distance for the first three modes. The plain lines represent the fitting using **Eq. (16)** with $E^{cell}$ as the only fitting parameter. The elastic modulus extracted from different modes increases with increasing the mode number (resonance frequency). The values of $E^{cell}$ for each mode (each resonance frequency) are shown in **FIG. 7**. As the frequency increases, the Young's modulus also increases.

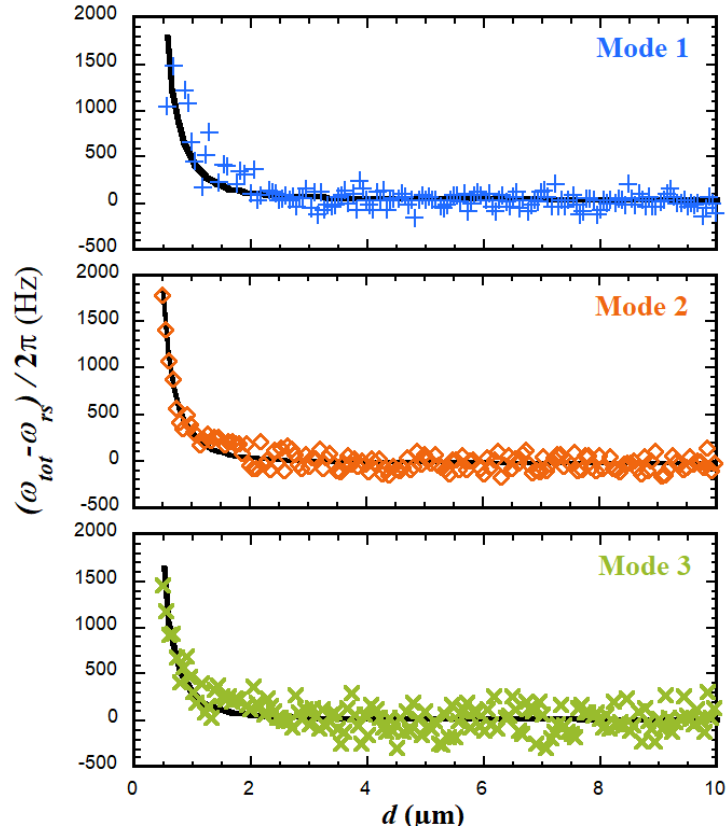

*FIG. 6 The frequency shift versus distance for the first three modes. The dots represent the value measured on the cell minus the value measured on the rigid sphere. The plain lines represent the fit using **Eq. (16)** for each mode.*

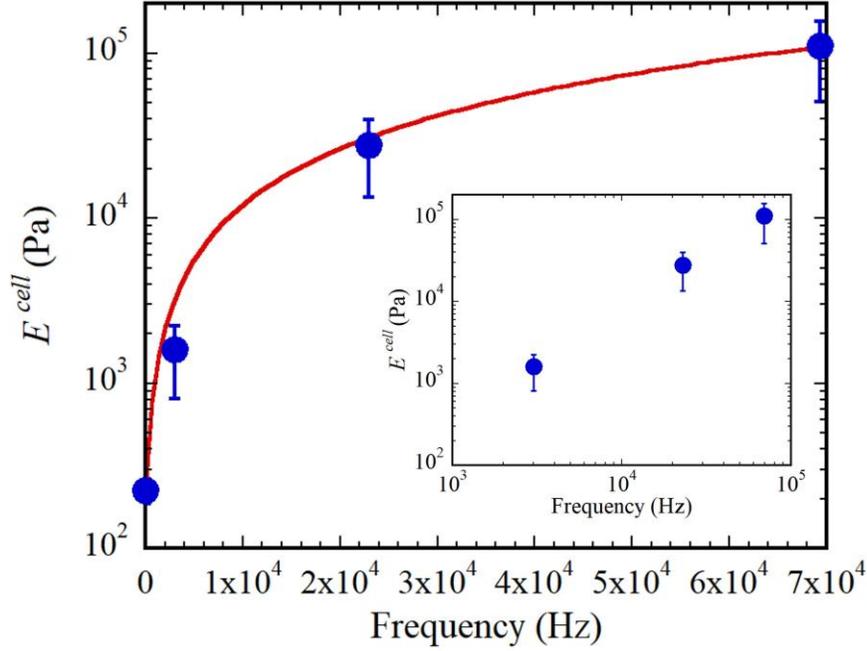

*FIG. 7 The fitted Young's modulus of the cell for the first three models is plotted against frequency. The inset shows the corresponding log-log plot. We also plot the static value extracted from the force-indentation curves as shown in **FIG. 8**. The red line represents the fit.*

The static Young's modulus $E_0^{cell}$ of the living cell can be detemined from the force-indentation curve. The AFM cantilever is positioned on the cell surface and then gently pressed against it with a very low piezo velocity (50 nm/s). To convert the measurements of deflection versus piezo displacement into force versus indentation, we follow these steps: 1) the static force is obtained by multiplying the measured static deflection of the cantilever by its spring constant, 2) the indentation depth ($\delta$) is obtained by subtracting the cantilever's deflection from the piezo displacement. This relationship between the force and the indentation depth is described by the Hertz model:

$$F = \frac{16}{9} E_0^{cell} R^{\frac{1}{2}} \delta^{\frac{3}{2}} \tag{17}$$

**FIG. 8** shows an example of the force-indentation curve measured on a living cell. The red solid line represents the fitted curve obtained using **Eq. (17)**, with $E_0^{cell}$ as the fitting parameter. From the fit, we calculate an average static elastic modulus of $E_0^{cell} = 225 \pm 40$ Pa.

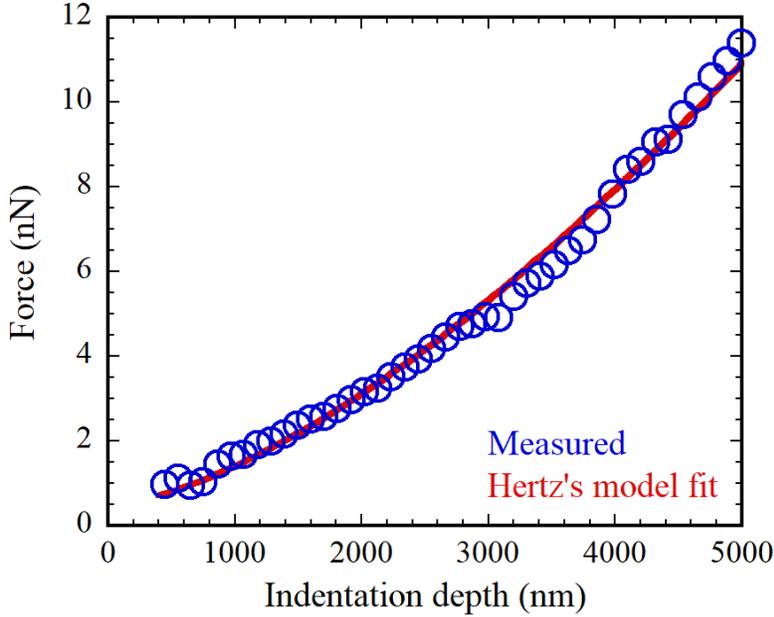

FIG. 8 *The force-indentation curve on the cell and the fitting using Hertz model (**Eq. (17)**).*

The extracted elastic modulus as a function of frequency can be fitted by a power law of the form: $E_0^{cell}(1 + af^n)$, where $E_0^{cell}$ and $f$ correspond to the static Young's modulus and frequency, respectively. The parameters $a$ and $n$ are the fitting coefficients. By fitting the data (red curve in **FIG. 7**), we obtain the values of $a = 0.00134$ and $n = 1.15$. The exponent obtained in our experiment is larger than the typical values reported in the literature, which are usually smaller than one [30-31]. In such experiments, the cell responses are measured by inducing large deformations of the cell at different frequencies. In contrast, in our non-contact measurement, the cell deformation is very small because we use nanometric cantilever vibrations (driven by thermal fluctuations). In our experiment we primarily investigated the surface response of the cell.

In conclusion we have used the nanoscale thermal fluctuations of an AFM cantilever to probe the visco-elastic properties of living cells. The elasto-hydrodynamic coupling between the cantilever fluctuations and the cell deformations was investigated by measuring the power spectral density (PSD) of the thermal fluctuations of the cantilever at different separation distances. The frequency shift due to the cell deformation extracted from the PSD was then fitted with an elasto-hydrodynamic model, combining the viscous drainage force and the cell deformation. The elastic modulus values of the living cell were determined for the first three resonance frequencies of the cantilever.

## ACKNOWLEDGMENTS

Hao Zhang acknowledges financial support from the China Scholarship Council. The authors acknowledge the French National Research Agency through the supporting grants EDDL (ANR-19-CE30-0012), AMARHEO (ANR-18-CE06-0009-01). The authors thank Anthony Bouter for providing the cells, and Antoine Allard for reading the manuscript and fruitful discussion.

## AUTHOR DECLARATIONS

Conflict of Interest

The author has no conflicts to disclose.

## DATA AVAILABILITY

The data that support the findings of this study are available from the corresponding authors upon reasonable request.